\documentclass[final]{svjour3b}

\usepackage{graphicx}
\usepackage{cite}
\usepackage{mathptmx}

\usepackage{comment}
\usepackage{color}

\newcommand{\Np}{N_+}\newcommand{\Nm}{N_-}\newcommand{\np}{n_+}\newcommand{\nm}{n_-}

\newcommand{\eps}{\epsilon}\newcommand{\epsP}{\epsilon_+}\newcommand{\epsM}{\epsilon_-}
\newcommand{\epsO}{\epsilon_0}\newcommand{\epsOP}{\epsilon_{0+}}\newcommand{\epsOM}{\epsilon_{0-}}
\newcommand{\KP}{K_+}\newcommand{\KM}{K_-}
\newcommand{\piP}{\pi_+}\newcommand{\piM}{\pi_-}
\newcommand{\piO}{\pi_0}\newcommand{\piOP}{\pi_{0+}}\newcommand{\piOM}{\pi_{0-}}\newcommand{\piOPM}{\pi_{0\pm}}
\newcommand{\vP}{v_+}\newcommand{\vM}{v_-}
\newcommand{\vB}{v_B}\newcommand{\vBP}{v_{B+}}\newcommand{\vBM}{v_{B-}}
\newcommand{\vF}{v_F}\newcommand{\vFP}{v_{F+}}\newcommand{\vFM}{v_{F-}}
\newcommand{\Fs}{F_s}\newcommand{\FsP}{F_{s+}}\newcommand{\FsM}{F_{s-}}

\newcommand{\Fd}{F_d}\newcommand{\FdP}{F_{d+}}\newcommand{\FdM}{F_{d-}}
\newcommand{\FC}{F_c}\newcommand{\vC}{v_c}
\newcommand{\Fp}{F_+}\newcommand{\Fm}{F_-}

\newcommand{\epsU}{\epsilon_{\rm U}}\newcommand{\piU}{\pi_{\rm U}}
\newcommand{\etxt}{{\rm e}}
\newcommand{\om}{\omega}

\newcommand{\piPM}{\pi_{+(-)}}
\newcommand{\vO}{v_0}\newcommand{\vOP}{v_{0+}}\newcommand{\vOM}{v_{0-}}

\newcommand{\Fig}[1]{Fig.~\ref{#1}}\newcommand{\Tab}[1]{Table~\ref{#1}}
\newcommand{\Figs}[2]{Figs.~\ref{#1} and \ref{#2}}
\newcommand{\Eq}[1]{equation~(\ref{#1})}
\newcommand{\Eqs}[2]{equations~(\ref{#1}) and (\ref{#2})}
\newcommand{\Sec}[1]{Section~\ref{#1}}

\newcommand{\App}[1]{Appendix \ref{#1}}

\newcommand{\ie}{i.e. }\newcommand{\eg}{e.g. }\newcommand{\resp}{resp. }

\newcommand{\invitro}{\textit{in vitro}}\newcommand{\invivo}{\textit{in vivo}}

\newcommand{\N}{(0)}\renewcommand{\P}{($+$)}\newcommand{\M}{($-$)}
\newcommand{\PM}{($-+$)}\newcommand{\PMN}{($-$0$+$)}\newcommand{\PN}{(0$+$)}\newcommand{\NM}{($-$0)}

\newcommand{\mum}{\mu{\rm m}}
\newcommand{\nanom}{\rm nm}
\newcommand{\pN}{{\rm pN}}
\newcommand{\s}{{\rm s}}

\begin{document}

\title{Motility states of molecular motors engaged in a stochastic tug-of-war}

\author{\begin{minipage}{\textwidth}Melanie J.I. M{\"u}ller \and  Stefan Klumpp \and Reinhard Lipowsky\end{minipage}}

\institute{
	Melanie J.I. M{\"u}ller and Reinhard Lipowsky \at Max Planck Institute of Colloids and Interfaces, Science Park Golm, 14424 Potsdam, Germany\\
	\email{mmueller@mpikg.mpg.de}
	\and Stefan Klumpp \at Center for Theoretical Biological Physics, University of California San Diego, La Jolla, CA 92093-0374, USA
	}

\date{}

\maketitle

\vspace*{-7em}
\begin{abstract}
Intracellular transport is mediated by molecular motors that pull cargos along cytoskeletal filaments. Many cargos move bidirectionally and are transported by two teams of motors which move into opposite directions along the filament.
We have recently introduced a stochastic tug-of-war model for this situation. This model describes the motion of the cargo as a Markov process on a  two-dimensional state space defined by the numbers of active plus and active minus motors. 
In spite of its simplicity, this tug-of-war model leads to a complex dependence of the cargo motility on the motor parameters. We present new numerical results for the dependence on the number of involved motors. 
In addition, we derive a simple and intuitive sharp maxima approximation, from which one obtains the cargo motility state from only four simple inequalities. This approach provides a fast and reliable method to determine the
cargo motility for a given experimental system.
\end{abstract}

\section{Introduction}

Molecular motors are cellular proteins which transform the free energy released from chemical reactions into mechanical work. Cytoskeletal motors walk in a directed fashion along cytoskeletal filaments and thereby transport cargo through the cell \cite{AlbertsWalter98}. Examples are kinesin and dynein, which walk along microtubule filaments, and certain myosins, which walk along actin filaments. Single motors of this kind have been studied extensively in recent years both experimentally \cite{Vale03} and theoretically \cite{LipowskyKlumpp05LifeIsMotion}. The filaments possess an intrinsic direction: they have one 'plus' and one 'minus' end. They  are essentially one-way roads for the motors. For example, cytoplasmic dynein walks to the microtubule minus end, while kinesin-1  walks to the microtubule plus end. Most cells have a unidirectional microtubule cytoskeleton \cite{LaneAllan98}: The microtubule minus ends are typically located near the cell center, while the plus ends point outwards towards the cell periphery. In special cells such as neurons with their long axonal protrusions, the microtubules form a unipolar parallel array with the minus ends pointing towards the cell center and the plus ends pointing towards the axon tip.

Although each motor walks only into one direction along such an isopolar filament network, many cellular cargos, like mitochondria, pigment granules and endosomes, are observed to move bidirectionally, reversing direction every few seconds \cite{Welte04}. Therefore, both plus-end and minus-end motors must be involved in the transport of a single cargo, and indeed both kinesin and dynein are found simultaneously on various cellular cargos
\cite{Gross04}.
We have recently proposed a stochastic tug-of-war model for this situation, in which the motors are coupled via the mechanical interaction with their common cargo \cite{MullerLipowsky08TugOfWar}. This model maps the motion of the cargo to a stochastic process on a finite two-dimensional state space defined by the numbers of active plus and active minus motors. Despite the simplicity of the model, the cargo exhibits a rather complex motility behavior. Depending on the motor parameters, the cargo motion exhibits several qualitatively different motility states, which correspond to 
qualitatively distinct steady state solutions of the corresponding Markov process. In particular, for biologically relevant parameter values, the cargo exhibits switching between fast plus and fast minus motion with or without interspersed pauses, as found in experiments \cite{Welte04,Gross04}. This fast bidirectional motion, which is usually associated with a coordination mechanism rather than with a tug-of-war, is obtained in our model via a dynamic instability, which leads to a high probability of having only one motor type bound at a given time. Therefore, the tug-of-war is a cooperative mechanism for bidirectional transport. In addition to the fast bidirectional motion, our tug-of-war model can explain many properties of bidirectional cellular cargo transport observed in experiments in a quantitative way \cite{MullerLipowsky08TugOfWar}. This implies that the signalling pathways that control intracellular transport may directly target the different motor molecules rather than an additional coordination complex \cite{MullerLipowsky08TugOfWar,WelteGross08}.

In our previous study in \cite{MullerLipowsky08TugOfWar}, we have emphasized the experimental relevance of our model and given a detailed comparison between theory and experiment. In this article, we  describe the modeling procedure in detail and present new numerical results for the parameter dependence of the motility states. The dependence on the motor numbers leads to the prediction that bidirectional cargo transport with pauses can only be accomplished by at least 4 motors on the cargo.
Furthermore, we describe a simple and intuitive approximation, which we call the `sharp maxima approximation' and which provides a reliable
description for the dependence of the different motility states on the single motor parameters. Using the analytical solution obtained for this approximation one can easily obtain the prediction of our model for the motion of a cargo carried by two motor species in a simple and transparent manner without performing tedious numerical calculations. The approximation also allows to calculate the effect of a change in the motor properties, which might be accomplished \eg by changing the concentration of ATP or salt, by adding regulatory molecules such as dynactin, or by mutation of the motors.

Cooperative effects in systems of many molecular motors have been studied extensively in the theoretical literature \cite{JulicherProst95,VilfanSchwabl99,Duke00,LipowskyNieuwenhuizen01PRLRandomWalksCompartments,BadoualProst02,KlumppLipowsky05PNASCargoTransport,GrillJulicher05,CampasJoanny06,GuntherKruse07}. Typically, these studies have considered large numbers of motors belonging to a single motor species as appropriate, e.g. for the modeling of muscles. In contrast, transport in cells, as mentioned, is often bidirectional because of the cooperation of (at least) two motor species and is typically based on rather small numbers of motors, in the range of 1-10 \cite{KlumppLipowsky05PNASCargoTransport,GrossShubeita07}.  Cargo transport over cellular length scales, both unidirectional and bidirectional, has been modeled in a coarse-grained way as movement of a particle with effective rates characterizing the speeds and run lengths of the cargo \cite{SmithSimmons01,LipowskyNieuwenhuizen01PRLRandomWalksCompartments,Maly02,PangarkarMitragotri05,LipowskyKlumpp05LifeIsMotion}.
While these models provide an appropriate description of cargo movements on large time and length scales, they cannot address the stochastic fluctuations of motor-filament binding and unbinding, the effect of the motors on each other, and the dependence of cargo transport on the molecular properties of the motors, which are addressed within the framework of our model.

This article is organized as follows. We introduce our stochastic tug-of-war model in \Sec{SecModel} and discuss its numerical solutions in \Sec{SecMotilityStates}. These two sections give a detailed description of the model introduced in \cite{MullerLipowsky08TugOfWar}, and summarize the main theoretical results reported in \cite{MullerLipowsky08TugOfWar}. In addition, new numerical results for varying motor numbers are shown.
In \Sec{SecSharpMax} we present a new analytic 'sharp maxima' approximation in order to explain this parameter dependence. This approximation allows to determine the motility states of a given system by using only four simple inequalities.

\section{Modeling}\label{SecModel}

\begin{figure}\centering
\includegraphics[width=7.5cm]{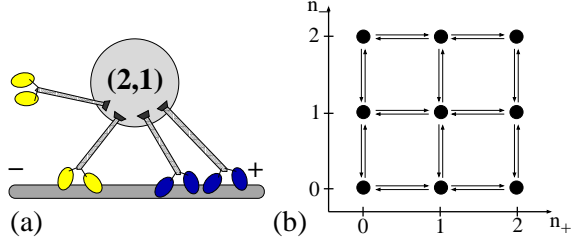}
\caption{\label{FigCartoonStateSpace}%
Cargo with a total number of $\Np=2$ plus (blue) and $\Nm=2$ minus (yellow) motors.
(a) The numbers $\np$ and $\nm$ of plus and minus motors that actually pull the cargo, which are $(\np,\nm)=(2,1)$ in the figure, fluctuate because the motors stochastically bind to and unbind from the filament.
(b) Two-dimensional state space of $(\Np+1)(\Nm+1)=9$ states labeled by $(\np,\nm)$.}
\end{figure}

We consider a cargo particle with a fixed number of $\Np$ plus and $\Nm$ minus motors. Typically these numbers will be in the range of 1 to 10 motors as observed for many cargos  \invivo\ \cite{KlumppLipowsky05PNASCargoTransport,GrossShubeita07}. For $\Np = 0$ or $\Nm= 0$, we recover the model for cooperative transport by a single motor species as studied in \cite{KlumppLipowsky05PNASCargoTransport}. Each of the motors can bind to, move along, and unbind from the filament. Because of the stochastic binding and unbinding, at each time only $\np$ plus and $\nm$ minus motors are bound to the filament, with $0\leq\np\leq\Np$ and $0\leq\nm\leq\Nm$, see \Fig{FigCartoonStateSpace}(a). Since only these bound motors can exert force on the cargo, the cargo motion is determined by the numbers $(\np,\nm)$. 
Unbinding or binding of a plus motor in the state $(\np,\nm)$ occurs with rate $\epsP(\np,\nm)$ or $\piP(\np,\nm)$ and changes this state to $(\np-1,\nm)$ or $(\np+1,\nm)$, respectively. Analogously, unbinding or binding of a minus motors occurs with rate $\epsM(\np,\nm)$ or $\piM(\np,\nm)$, respectively, This leads to a 'random walk' on the state space shown in \Fig{FigCartoonStateSpace}(b). This random walk or Markov process is described by a Master equation for the probability $p(\np,\nm,t)$ to have $\np$ active plus and $\nm$ active minus motors at time $t$. This Master equation has the form
\begin{eqnarray}\label{EqMasterEq}
\frac{\partial}{\partial t} p(\np,\nm,t) &=& p(\np+1,\nm,t)\,\epsP(\np+1,\nm) + p(\np,\nm+1,t)\,\epsM(\np,\nm+1)\nonumber\\
 &&\hspace{-1em}+\;p(\np-1,\nm,t)\,\piP(\np-1,\nm) + p(\np,\nm-1,t)\,\piM(\np,\nm-1)\nonumber\\
 &&\hspace{-1em} - p(\np,\nm,t)\, \left[\piP(\np,\nm)+\piM(\np,\nm)+\epsP(\np,\nm)+\epsM(\np,\nm)\right].
\end{eqnarray} 
In order to obtain the rates $\eps_{+(-)}$ and $\piPM$ for unbinding and binding, respectively, of one motor, we first establish a model for a single motor, which is based on the transport properties of molecular motors as measured in single-molecule experiments. We then derive the effective cargo rates that enter the Master \Eq{EqMasterEq} by assuming that the motors feel each other only because they act on their common cargo, as explained below.

\begin{table}
\begin{tabular}{l|c|l|l}
	Parameter 		& symbol 	& kinesin-1							& cytoplasmic dynein\\\hline 
	stall force 		& $\Fs$ 	& 6 pN		 \hfill\cite{SchnitzerBlock00,SvobodaBlock94}
							& 7 pN ('strong')		\hfill\cite{TobaHiguchi06}\\
					&		&& 1.1 pN ('weak')\,\hfill\cite{MallikGross05,WelteWieschaus98}\\
	detachment force 	& $\Fd$ 	& 3 pN		 \hfill\cite{SchnitzerBlock00}			& ?\\
	unbinding rate 	& $\epsO$ & 1/s		 \hfill\cite{SchnitzerBlock00,ValeYanagida96}& 0.25/s \hfill\cite{KingSchroer00,ReckPetersonVale06}\\
	binding rate 		& $\piO$ 	& 5/s		 \hfill\cite{LeducProst04,BeegLipowsky08}    & 1.5/s \hfill\cite{CarterCross05,ValeYanagida96}\\
	forward velocity	& $\vF$ 	& 1 $\mu$m/s\,	 \hfill\cite{CarterCross05,ValeYanagida96}   & 1 $\mu$m/s \hfill\cite{KingSchroer00,TobaHiguchi06} \\
	superstall velocity amplitude	& $\vB$	& 6 nm/s		 \hfill\cite{CarterCross05}				& ?
\end{tabular}
\caption{\label{KinesinParameters}%
Single motor parameters of our model, and values for kinesin-1 and cytoplasmic dynein, taken from the cited references. A question mark indicates that this parameter is not available.}
\end{table}

\subsection{Model for a single motor}

A single motor can bind to a filament with rate $\pi$, walk along it with velocity $v$, and unbind from it with rate $\eps$. In our motor tug-of-war, opposing motors exert force on each other, so that the load-dependence of these parameters is crucial. For some molecular motors such as kinesin-1, this load-dependence has been measured in single molecule experiments. The motor velocity $v$ decreases with increasing load force from its zero-load value $\vF$ to zero at the so-called stall force $\Fs$. This decrease is approximately linear, and given by
\begin{eqnarray}\label{EqSingleVF}
	v(F) = \vF\left(1-F/\Fs\right) \quad\mbox{for } 0\leq F\leq\Fs,
\end{eqnarray}
as has been observed for kinesin-1 \cite{SvobodaBlock94,NishiyamaYanagida02,CarterCross05}, for kinesin-3 \cite{TomishigeVale02} and cytoplasmic dynein \cite{TobaHiguchi06}. Here we use the convention that the load force $F$ is positive if it acts opposite to the motor's forward direction. For superstall load forces $F>\Fs$, the motor can walk backwards, but only slowly, as has been shown for kinesin-1 \cite{NishiyamaYanagida02,CarterCross05} and cytoplasmic dynein \cite{KojimaOiwa02,MallikGross04Dyn}. 
In this regime, the functional form of the motor force-velocity relation is unclear; for simplicity, we use a linear form
\begin{eqnarray}\label{EqSingleVB}
	v(F) = \vB\left(1-F/\Fs\right) \quad\mbox{for }  F\geq\Fs.
\end{eqnarray}
The superstall velocity amplitude $\vB$ characterizes the slope $\vB/\Fs$ of the force-velocity relation in the superstall regime.
Our results are essentially independent of the exact form of the motor force-velocity-curve as long as it decreases monotonically with external force and exhibits slow backward motion. 
The unbinding $\eps$ rate of the motor, which equals $\epsO$ at zero load force, increases exponentially with the load force as 
\begin{eqnarray}\label{EqSingleEps}
	\eps(F) = \epsO\,\exp\left[F/\Fd\right],
\end{eqnarray}
which defines the detachment force $\Fd$.
Such a functional form has been measured for kinesin-1 \cite{SvobodaBlock94}, and also follows theoretically from Kramers or Bell theory \cite{Bell78}.
The force-dependence of the binding rate is difficult to assess experimentally. However, it is expected to depend only weakly on the load force because an unbound motor relaxes and then binds from its relaxed state. We therefore take the binding rate equal to
\begin{eqnarray}\label{EqSinglePi}
	\pi(F) = \piO
\end{eqnarray}
for all load forces $F$.

All single-motor parameters have been measured for the motor kinesin-1, see \Tab{KinesinParameters}. Other motors have parameters of similar orders of magnitude, but their parameters are not as well established. For cytoplasmic dynein, for example, no measurements for the detachment force and for the superstall velocity amplitude are available, and different labs have reported different values for the stall force, see  \Tab{KinesinParameters}. 
In this work, we use the kinesin-1 parameters for the missing dynein parameters, and we call dyneins with stall force 7 pN 'strong' and with 1.1 pN 'weak' dyneins.

\subsection{Effective rates for the cargo}

We assume that the motors on the cargo act independently and feel each other only because 
(i) opposing motors act as load, and 
(ii) same-directional motors share this load. 
If each plus motor feels the load $\Fp$ (and generates the force $-\Fp$) and each minus motor feels the load $-\Fm$ (and generates the force $\Fm$), this means that the force balance on a cargo pulled by $\np$ plus and $\nm$ minus motors is
\begin{eqnarray}\label{EqForceBalance}
    \np\Fp = -\nm\Fm \equiv\FC(\np,\nm),
\end{eqnarray}
where $\FC$ denotes the force which acts on the team of the plus as well as on the team of the minus motors.
Here, the sign of the force is chosen positive if it is a load on the plus motors, \ie if it points into the minus direction. If only one motor type is bound, \ie if $\np=0$ or $\nm=0$, then $\Fp=\Fm=\FC=0$. The force balance as given by (\ref{EqForceBalance}) represents Newton's third law: each motor feels the same force as it exerts.
A single bound plus motor thus feels the force $\Fp=\FC/\np$. Using this in the single motor unbinding rate (\ref{EqSingleEps}), this implies that the effective rate for the unbinding of one plus motor in the state $(\np,\nm)$ is
\begin{eqnarray}\label{EqCargoEpsP}
	\epsP(\np,\nm) = \np\,\epsOP\,\exp\left[\FC(\np,\nm)/(\np\FdP)\right].
\end{eqnarray}
Here and below, the index '$+$' labels plus motor properties, e.g. $\FdP$ is the detachment force of a single plus motor.
The prefactor $\np$ in \Eq{EqCargoEpsP} describes that there are $\np$ bound plus motors available for unbinding.
Analogously, the effective rate for the binding of one plus motor in the state $(\np,\nm)$, where there are $\Np-\np$ unbound plus motors, is, with \Eq{EqSinglePi},
\begin{eqnarray}\label{EqCargoPiP}
	\piP(\np,\nm) = (\Np-\np)\,\piOP.
\end{eqnarray}
Analogous expressions hold for the unbinding and binding rate of a minus motor with the parameters indexed by '$-$'.

The cargo force is determined by the condition that both plus motors, which experience the force $\Fp$, and minus motors, which experience the force $-\Fm$, move with the same velocity $\vC$ as given by
\begin{eqnarray}\label{EqVelocityBalance}
    \vC(\np,\nm) = \vP(\Fp) = -\vM(-\Fm).
\end{eqnarray}
Here, the sign of the velocity is taken to be positive in the plus direction and negative in the minus direction. 
The velocity balance relation (\ref{EqVelocityBalance}) assumes that all motors walk with the same velocity. This can be justified by considering what happens when this steady state is disturbed, \eg by the binding of a minus motor while the cargo is steadily moving into the plus direction. The 'new' motor has bound in a relaxed state and feels a low load force. It therefore steps forward, gets stretched and burdens itself with part of the load force imposed by the plus motors, taking away part of the load force of the other minus motors. It does so until all minus motors feel roughly the same load force.
This justifies our mean field treatment that each plus or minus motor feels on average the \textit{same} force $\Fp$ or $-\Fm$. The new force balance depends on whether the plus or minus motors are 'stronger'. If the plus motors together can produce a larger force than the minus motors, \ie if $\np\FsP>\nm\FsM$ (with the 'new' number $\nm$ of bound minus motors), then the new minus motor runs forward until it feels a superstall load force and then starts to run backwards. The total force on the cargo increases, and all motors slow down until a new force balance is achieved.
If the minus motors can produce a larger force than the plus motors, \ie if $\nm\FsM>\np\FsP$, the minus motors 'take over': the new minus motor runs forward and takes away more and more load force from the other minus motors until they all feel a substall load force $-\Fm<\FsM$. The plus motors, on the contrary, start to move backward as soon as they feel a superstall load force $\Fp>\FsP$. The forward and backward stepping of motors thus provides a relaxation mechanism which establishes the force and velocity balances as provided by \Eqs{EqForceBalance}{EqVelocityBalance}. 

The force and velocity balances as given by \Eq{EqForceBalance} and (\ref{EqVelocityBalance}) lead to the cargo force 
\begin{eqnarray}
	\FC(\np,\nm) &=& \lambda(\np,\nm)\, \np\FsP + \left[1-\lambda(\np,\nm)\right]\,\nm\FsM, 	\label{EqFC}
\end{eqnarray}
with  
\begin{eqnarray}
\lambda(\np,\nm)=1/\left[1+(\np\FsP\vOM)/(\nm\FsM\vOP)\right],
\end{eqnarray}
and to the cargo velocity
\begin{eqnarray}
	\vC(\np,\nm) &=& \frac{\np\FsP - \nm\FsM}{\np\FsP/\vOP+\nm\FsM/\vOM}. \label{EqVC}
\end{eqnarray}
Here the single motor velocity parameter $\vOP$ is equal to the plus motor forward velocity $\vFP$ or superstall velocity amplitude $\vBP$, depending on whether the plus motors move forward, \ie $\vC>0$, or backward, \ie $\vC<0$, respectively. Analogously, the single motor velocity parameter $\vOM$ is equal to $\vFM$ for $\vC<0$ and to $\vBM$ for $\vC>0$.
It follows from \Eq{EqVC} that the cargo direction is determined by the intuitive 'majority rule'
\begin{eqnarray}\label{EqMajorityRule}
    \vC>0 \quad\quad\mbox{ for }\; \np\FsP &>& \nm\FsM,  \quad\quad\mbox{\ie plus motors 'win'} \nonumber\\
    \vC<0 \quad\quad\mbox{ for }\; \np\FsP &<& \nm\FsM,  \quad\quad\mbox{\ie minus motors 'win'} \\
    \vC=0 \quad\quad\mbox{ for }\; \np\FsP &=& \nm\FsM,  \quad\quad\mbox{\ie 'tie'.}\nonumber
\end{eqnarray}
Since $0\leq\lambda\leq 1$, the cargo force as in \Eq{EqFC} is a convex combination of the maximal plus and minus motor stall forces $\np\FsP$ and $\nm\FsM$. Therefore,  if for example the plus motors win, these motors walk under substall and the minus motors under superstall load force, \ie $\nm\FsM < \FC(\np,\nm) < \np\FsP$, which is consistent with the single motor force-velocity relation \Eqs{EqSingleVF}{EqSingleVB}.
 
The cargo force given by \Eq{EqFC} determines the plus motor rates (\ref{EqCargoEpsP}) and (\ref{EqCargoPiP}) and the corresponding minus motor rates for the Master \Eq{EqMasterEq}. However, one has to take care of reflecting boundary conditions, which ensure that the numbers $(\np,\nm)$ of active motors stay within the intervals  $0\leq\np\leq\Np$ and $0\leq\nm\leq\Nm$. 

Experiments \invivo\ usually monitor cargos which have been walking along a filament for some (unknown) time. This has two implications for our calculations: 
First, we are interested in long-time properties of cargo motion, \ie in the time-independent steady state probability $p(\np,\nm)$ to have $\np$ active plus and $\nm$ active minus motors, which is obtained by setting the time-derivative in the Master \Eq{EqMasterEq} equal to zero.
Second, we are interested in probabilities that are conditioned on the cargo being bound to the filament. 
We can separate cargo unbinding into two steps: (i) the last motor unbinds from the filament so that the cargo is in the state $(\np,\nm)=(0,0)$ but still close to the filament, and (ii) the cargo diffuses away from the filament and then undergoes free diffusion in the surrounding solution. We call this diffusing state $U$. The rates for cargo unbinding into and rebinding from the state $U$ depend on the geometry of the system and the viscosity of the surrounding solution. In \App{AppCargoUnbinding} we show that the probability $p(\np,\nm)$ which solves the time-independent Master \Eq{EqMasterEq} is in fact the steady-state probability conditioned on the cargo being bound to the filament, \ie not to be in state $U$, and that this probability is independent of the rates that connect the state $(0,0)$ with state $U$. We can therefore ignore cargo unbinding and diffusion. Furthermore, a cargo in the state $(0,0)$ is close to the filament. Binding of plus and minus motors from this state is then described by the intrinsic motor rates $\piOP$ and $\piOM$, respectively, which justifies our choice of the binding rates in (\ref{EqCargoPiP}) also for the state $(\np,\nm)=(0,0)$.

We solve the Master \Eq{EqMasterEq} for the steady state by determining the eigenvector of the associated transition matrix with eigenvalue zero. In addition, we simulate individual cargo trajectories by using the Gillespie algorithm \cite{Gillespie76} for the binding/unbinding dynamics as given by \Eqs{EqCargoEpsP}{EqCargoPiP} and let the cargo move with velocity $\vC$ in the intervals between (un-)binding events. 

The Master \Eq{EqMasterEq} describes a two-dimensional random walk on the network shown in \Fig{FigCartoonStateSpace}(b). The steady-state solution of the Master \Eq{EqMasterEq} is a nonequilibrium steady state, as can be seen by considering the ratio of the product of forward and the product of backward rates of the cycles of the network of \Fig{FigCartoonStateSpace}(b). This product is not equal to 1 for all cycles, as would be required for an equilibrium state \cite{LipowskyKlumpp05LifeIsMotion,LiepeltLipowsky07SteadyState}.
The rates of the Master equation are nonlinear in the state space variables $(\np,\nm)$ because of the nonlinear force-dependence of the unbinding rates (\ref{EqCargoEpsP}), which leads to a cooperative effect, namely the unbinding cascade described in the next section. The emergence of cooperative behavior arising from the nonlinear force dependence of the unbinding rate has also been proposed as an explanation for collective effects in muscles \cite{Duke00,GuntherKruse07} and mitotic spindle oscillations \cite{GrillJulicher05}.

\section{Motility states}\label{SecMotilityStates}

Depending on the number and the parameters of the motors on the cargo, the cargo exhibits qualitatively different types of motions which we call 'motility states'. We define the motility states via the number and locations of the maxima of the motor number probability $p(\np,\nm)$ in the steady state. These maxima characterize the cargo motion because the cargo spends most of its time in these maxima $(\np,\nm)$. With respect to cargo motion, there are three different types of maxima: a maximum at $(\np,\nm)$ with $\np>0$, $\nm=0$ corresponds to fast plus motion with cargo velocity $\vC=\vFP$, a maximum at $(\np,\nm)$ with $\np=0$, $\nm>0$ corresponds to fast minus motion with cargo velocity $\vC=\vFM$, and a maximum at $(\np,\nm)$ with $\np,\;\nm>0$ corresponds to small cargo velocity $\vC$. The latter can be seen from \Eq{EqVC} by considering the fact that biological motors have a small superstall velocity amplitude $\vB\ll\vF$. Expanding the cargo velocity $\vC(\np,\nm)$ in, \eg for winning plus motors with $\np\FsP>\nm\FsM$, $\vBM/\vFP$, leads to $\vC\approx\vBM\left[\np\FsP/(\nm\FsM)-1\right]$, which is small for small $\vB$. Intuitively, the losing minus motors have to walk backwards, which they do only slowly, so that cargo motion is slowed down correspondingly.

\begin{figure}\centering
\includegraphics[width=9.5cm]{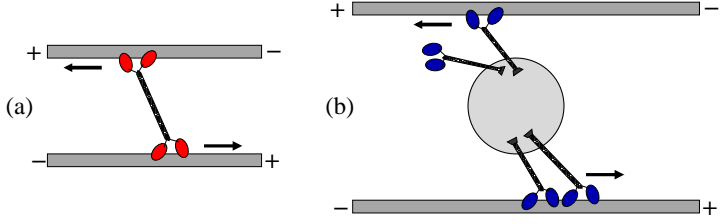}
\caption{\label{FigAntiparallelMTs}%
Possible realizations of the symmetric tug-of-war. 
(a) The tetrameric motor Eg5, walking along antiparallel microtubules in the mitotic spindle, corresponds to a symmetric tug-of-war of $\Np=1$ 'plus' motor (the upper half of the motor) and $\Nm=1$ 'minus' motor (the lower half of the motor).
(b) A large cargo carried by four motors of the same type along two antiparallel microtubules corresponds to a symmetric tug-of-war of $\Np=2$ 'plus' (upper) and $\Nm=2$ 'minus' (lower) motors.}
\end{figure}

\subsection{Symmetric tug-of-war}

It is instructive to consider the 'symmetric' tug-of-war for the same number $N=\Np=\Nm$ of plus and minus motors with identical single motor parameters apart from their forward directions. 
This case is theoretically appealing because of its simplicity and intuitive appeal. In addition, the effects of the motor tug-of-war are most pronounced in the symmetric case because the motors are 'equally strong'. 
Moreover, this symmetric situation may not be too unrealistic for transport \invivo\ where plus end and minus end transport are often found to exhibit astonishingly similar properties \cite{WelteWieschaus98,GrossGelfand02,PangarkarMitragotri05}. Furthermore, the most relevant plus and minus motors, kinesin-1 and cytoplasmic dynein, exhibit similar properties in terms of their processivities \cite{KingSchroer00,SchnitzerBlock00} and their force-velocity-curves \cite{TobaHiguchi06}, although the latter is still controversial, see \eg \cite{GennerichVale07}. The symmetric tug-of-war can also be realized by using only one type of motor but antiparallel filaments, see \Fig{FigAntiparallelMTs}. 
The tetrameric kinesin-5 motor Eg5 walks along antiparallel microtubules in the mitotic spindle during cell division. This corresponds to a symmetric tug-of-war of $\Np=1$ 'plus' motor, corresponding to one dimer, and $\Nm=1$ 'minus' motor, corresponding to the other dimer of the Eg5 tetramer, see \Fig{FigAntiparallelMTs}(a). In a possible \invitro\ experiment, a large cargo is transported by several motors of one type along antiparallel microtubules, see \Fig{FigAntiparallelMTs}(b). This again corresponds to a symmetric tug-of-war of 'plus' and 'minus' motors, where the assignment of a motor to the 'plus' or 'minus' motor type depends on the microtubule to which it attaches.

In the symmetric situation, the indices for plus and minus motors for the single motor parameters can be omitted, i.~e.: 
\begin{equation}
\piO\equiv\piOP=\piOM,\, \epsO\equiv\epsOP=\epsOM,\,\Fs\equiv\FsP=\FsM,\,\Fd\equiv\FdP=\FdM,\,\vO\equiv\vOP=\vOM
\end{equation}
In this symmetric case, inspection of the Master equation \Eq{EqMasterEq} and its rates (\ref{EqCargoEpsP}) and (\ref{EqCargoPiP}) shows that the steady state probability $p(\np,\nm)$ only depends on four dimensionless parameters: 
\begin{eqnarray}
\mbox{   the motor number } N\equiv\Np=\Nm,\mbox{ the desorption constant } K\equiv\epsO/\piO,\nonumber\\
\mbox{the force ratio } f\equiv\Fs/\Fd, \mbox{ and the velocity ratio } \nu\equiv\vB/\vF. \label{EqSymNondimParas}
\end{eqnarray}
As most motors walk backwards rather slowly, the velocity ratio is very small for most motors, e.~g. $\nu=0.006$ for kinesin-1. Because of the small value of $\nu$, the results are rather insensitive to its precise value  and very close to the results for no backward motion, i.~e. $\nu=0$. For fixed numbers $N$ of motors on the cargo, the relevant parameters are therefore the force ratio $f$ and the desorption constant $K$. The latter parameter describes the binding affinity of the motors; for processive motors like kinesin or dynein it is smaller than 1.
If the force ratio $f=\Fs/\Fd>1$, the force $\Fs$ that the motor exerts exceeds the force $\Fd$ that it can sustain. 
This ratio is most important for the qualitative properties of the steady state solution $p(\np,\nm)$ of the Master \Eq{EqMasterEq}. As described in \cite{MullerLipowsky08TugOfWar}, there are three types of solutions, which we call motility states, and which differ in the number and locations of the maxima of $p(\np,\nm)$, see \Fig{FigMotCharSymN5} and \Fig{FigMotDiagSymN5}:

\begin{figure}\centering
\includegraphics[width=11cm]{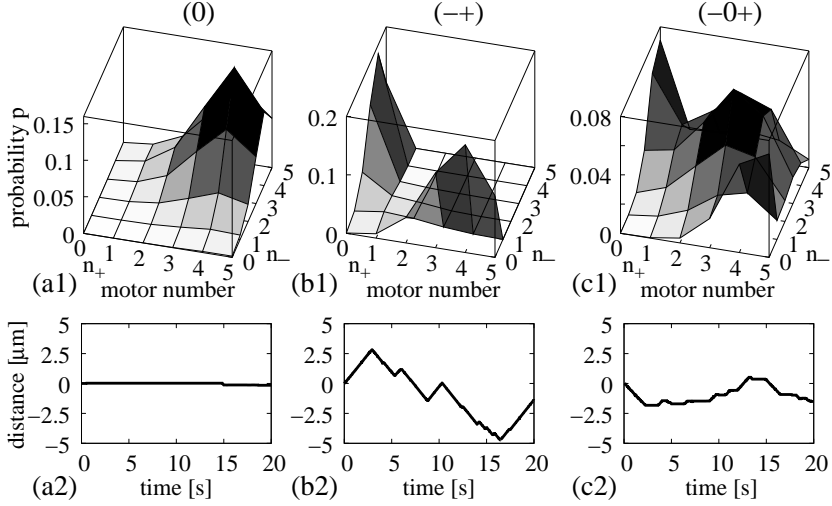}
\caption{\label{FigMotCharSymN5}%
The motility states for the symmetric tug-of-war of $N=\Np=\Nm=5$ plus and minus motors are characterized by qualitatively different motor number probability distributions (top row) and cargo trajectories (bottom row), as described in the text. Motor parameters are for kinesin as given in \Tab{KinesinParameters} except for 
(a) $\Fs=0.4\Fd$, $\epsO=0.2\piO$,
(b) $\epsO=0.35\piO$, 
(c) $\Fs=1.0\Fd$, $\epsO=0.2\piO$.
In all cases the superstall velocity amplitude is scaled as $\vB=1\,\nanom/\s\cdot\Fs/\pN$.
}
\end{figure}

\textbf{\N\ No motion}: For 'weak motors' with small stall to detachment force ratios $f=\Fs/\Fd$, the motors are hardly affected by the presence of the opposing motors and bind / unbind as if they were alone. The probability distribution $p(\np,\nm)$ has a single symmetric maximum at a state with $\np=\nm$, see \Fig{FigMotCharSymN5}(a1), corresponding to a state with zero velocity. The cargo therefore shows only small fluctuations around its starting position, see \Fig{FigMotCharSymN5}(a2). 

\textbf{\PM\ Fast bidirectional motion}: For 'strong motors' with large force ratio $f$, the motors feel the opposing motors strongly, and spontaneous symmetry breaking occurs. The probability distribution $p(\np,\nm)$ develops two maxima at $(\np,\nm)=(n,0)$ and $(0,n)$, see \Fig{FigMotCharSymN5}(b1), which correspond to fast motion into the plus and minus direction, respectively. The cargo stochastically switches between these two maxima and thereby between steady plus and minus motion, see \Fig{FigMotCharSymN5}(b2). The emergence of these two maxima can be understood in the following way: When, \eg because of a stochastic fluctuation, more plus than minus motors are bound to the filament ($\np>\nm$), every plus motor experiences the load force $\FC/\np$, while every minus motor experiences the larger load force $\FC/\nm$. Since the unbinding rate increases exponentially with increasing load force according to \Eq{EqSingleEps}, minus motors are more likely to unbind from the filament than plus motors. After the unbinding of a minus motor, the remaining minus motors experience an even larger load force and are even more likely to unbind. As a consequence, an unbinding cascade of minus motors happens until no minus motor remains bound. 
A prerequisite for this unbinding cascade is that the motors can exert a sufficiently large force to pull off opposing motors from the filament, \ie the stall force $\Fs$ has to be comparable to or larger than the detachment force $\Fd$. For small force ratios $f=\Fs/\Fd$, the pulling force has only a small effect on motor unbinding, so that no instability occurs and the cargo exhibits the blocked motility state \N. For large motor force ratio, the transient predominance of one motor type is amplified and stabilized by the described dynamic instability, and the cargo spends most of the time in states with only one motor type bound. 

\textbf{\PMN\ Fast bidirectional motion with pauses}: For intermediate force ratios $f$, the motor number probability $p(\np,\nm)$ displays three maxima, see \Fig{FigMotCharSymN5}(c1), a symmetric one with $\np=\nm$ corresponding to no motion as in \N, one with $\nm=0$ and one with $\np=0$, corresponding to fast plus and minus motion as in \PM. The cargo trajectory therefore exhibits bidirectional motion interrupted by pauses, see \Fig{FigMotCharSymN5}(c2).

\begin{figure}\centering
\includegraphics[width=9cm]{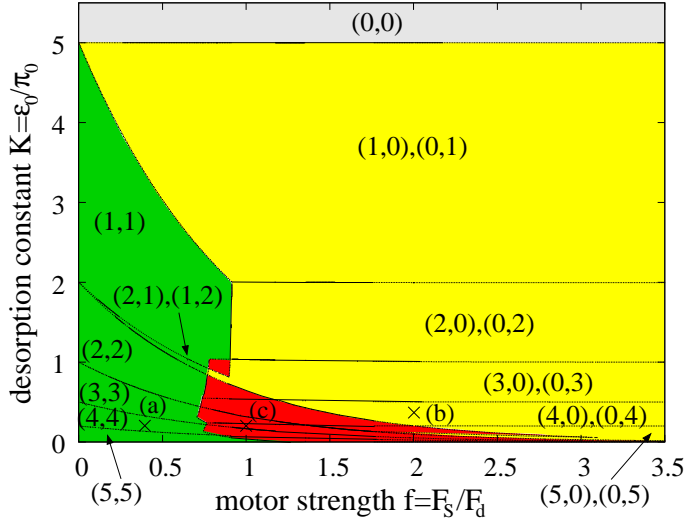}
\caption{\label{FigMotDiagSymN5}%
The motility diagram for the symmetric tug-of-war of $N=\Np=\Nm=5$ plus and minus motors displays the dependence of the cargo motility on the motor force ratio $f=\Fs/\Fd$ and the filament desorption constant $K=\epsO/\piO$. The lines separate regions with different locations $(\np,\nm)$ or numbers of maxima. The colors separate the different motility states of no motion \N\ (green), fast bidirectional motion \PM\ (yellow) and fast bidirectional motion with pauses (red). When the maximum is at $(0,0)$, the cargo is unbound (grey). 
Parameters are for kinesin-1 as in \Tab{KinesinParameters}, except for the stall force $\Fs$ and unbinding rate $\epsO$, which are varied. In addition, the superstall velocity amplitude is scaled as $\vB=1\,\nanom/\s\cdot\Fs/\pN$ in order to keep the backward slope $\vB/\Fs$ of the superstall force-velocity \Eq{EqSingleVB} constant.
The crosses labeled (a), (b) and (c) correspond to the parameter sets of \Fig{FigMotCharSymN5}.
}
\end{figure}

\Fig{FigMotDiagSymN5} shows the classification of motility states for the tug-of-war of $N=\Np=\Nm=5$ kinesin-like motors with respect to the relevant motor parameters $f=\Fs/\Fd$ and $K=\epsO/\piO$. As discussed above, the cargo is in one of the three motility states \N, \PM\ or \PMN. The lines in the motility diagram of \Fig{FigMotDiagSymN5} separate regions in which the maxima of the motor number probability distribution $p(\np,\nm)$ are located at different motor number states $(\np,\nm)$. The colors separate regions with different motility states. The detailed calculation procedure is as follows: The single-motor parameters are taken to be equal to the kinesin-1 values as given in \Tab{KinesinParameters}, except for the plus and minus motor unbinding rates $\epsO$ and stall forces $\Fs$. The parameter space $(\Fs,\epsO)$ is then explored systematically. All other parameters, \ie $\Fd$, $\piO$, $\vF$ and $\vB/\Fs$, are kept constant. Note that rather than the superstall velocity amplitude $\vB$, the slope $\vB/\Fs$ of the superstall force-velocity-curve (\ref{EqSingleVB}) is kept constant.\footnote{The need for emphasis of this point was kindly pointed out to us by Harry W. Schroeder III of the Goldman lab.} For each point $(\Fs,\epsO)$, the maxima of the motor number probability $p(\np,\nm)$ is calculated as the eigenvector with zero eigenvalue of the transition matrix of the Master \Eq{EqMasterEq}.  When the maxima between two scanned points change, we zoom in between these points in order to determine the transition point more accurately. The lines shown in \Fig{FigMotDiagSymN5} consist of these points.

\begin{figure}\centering
\includegraphics[width=\textwidth]{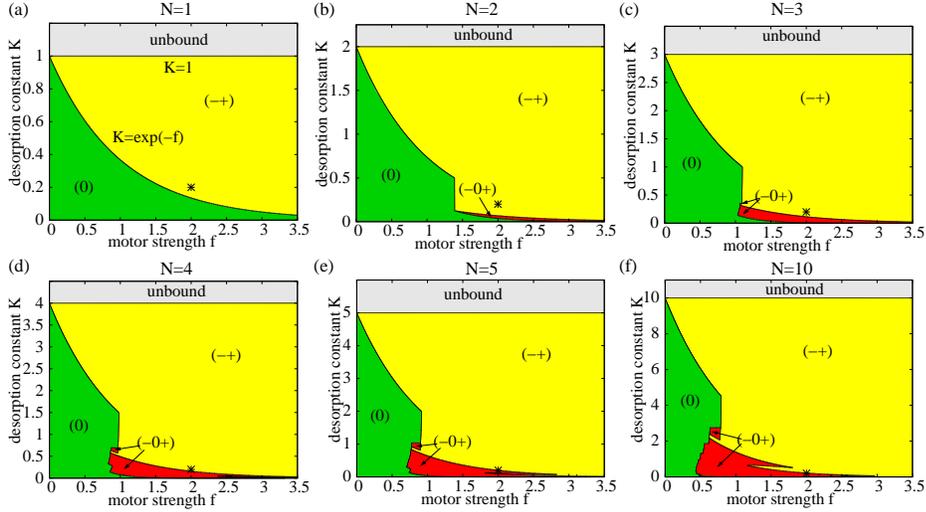}
\caption{\label{FigMotDiagSymN}%
Motility diagrams for the symmetric tug-of-war of $N=\Np=\Nm$ plus and minus motors. Color code and motor parameters as in \Fig{FigMotDiagSymN5}.
(e) is identical to \Fig{FigMotDiagSymN5}, but only the transition lines that separate different motility states are shown.
The crosses correspond to the full kinesin-1 parameter set of \Tab{KinesinParameters}.
The transition lines for $N=1$  can be determined analytically as given in (a).}
\end{figure}

For large desorption constants $K$, the motors have a low affinity to the filament; therefore the number of bound motors at the maxima of $p(\np,\nm)$ in \Fig{FigMotDiagSymN5} is low for high $K$. For very high desorption constants $K$, the maximum is at $(n,n)=(0,0)$, and the cargo is 'unbound'. 
For small force ratios $f$, the probability distribution $p(\np,\nm)$ has a single maximum at a no-motion state with an equal number of plus and minus motors bound at $(n,n)$ with $1\leq n\leq 5$, and the cargo is in the no-motion motility state \N\ (green). The two neardiagonal maxima at (1,2) and (2,1) are also counted as a single diagonal maximum, which in a continuous state space would be at $(x,x)$ with $1<x<2$.
For large force ratios $f$, the motors can generate forces large enough to rip off opposing motors since the stall force is large compared to the detachment force. This leads to the unbinding cascade described above, and the motor number probability has two maxima, one at a state $(n,0)$ with only active plus and one at a state $(0,n)$ with with only active minus motors, with $1\leq n\leq N$. In the latter situation, the cargo is in the \PM\ motility state (yellow). For intermediate values of $f$, both types of maxima coexist, and the cargo is in the \PMN\ motility state with three maxima of the motor number probability $p(\np,\nm)$ (red). 

\Fig{FigMotDiagSymN} shows the motility diagrams for varying motor numbers $N=\Np=\Nm$, where only the transition lines between the different motility states are shown.  
If the cargo is carried by only $\Np=1$ plus and $\Nm=1$ minus motor, the motility diagram can be determined analytically. The motor number probability $p(\np,\nm)$ has its maxima either at $(1,1)$ for $K<\exp(-f)$ (motility state \N), at $(0,0)$ for $K>1$ (unbound), or at $(1,0)$, $(0,1)$ otherwise (motility state \PM). The region with the three-maxima motility state \PMN\ only appears when the number $N=\Np=\Nm$ of motors on the cargo is larger or equal to 2, and increases for larger motor numbers, see \Fig{FigMotDiagSymN}. 

The crosses in the motility diagrams of \Fig{FigMotDiagSymN} correspond to the full set of kinesin-1 parameters as given in \Tab{KinesinParameters}. As other molecular motors have parameters of a similar order of magnitude, all three motility states \N, \PM\ and \PMN\ are within biological parameter range. Therefore, the cell can use tuning of the motor parameters in order to drastically change the motion of the cargo. 

The locations of the transition lines in the motility diagrams of \Figs{FigMotDiagSymN5}{FigMotDiagSymN} are determined by the balance of binding and unbinding of single motors under the load force generated by the opposing motors. As all unbinding rates are proportional to $\epsO$, and all binding rates are proportional to $\piO$, this means that the transition lines should scale with $K=\epsO/\piO$. Indeed, when the desorption constant $K$ is scaled by the motor number $N$, the diagrams of \Fig{FigMotDiagSymN}(b)-(f) almost overlay. 
As a specific example, cargo unbinding occurs when the rate of unbinding of the last bound motor, $\epsO$, becomes larger than the rate for binding of one motor in the unbound state, $N\piO$, \ie when $K=\epsO/\piO>N$. This is indeed the case in \Fig{FigMotDiagSymN} for all $N$. For the other transition lines, this type of reasoning is not exact but leads to a good approximation as we will be shown in \Sec{SecSharpMax}. 

\begin{figure}\centering
\includegraphics[width=11cm]{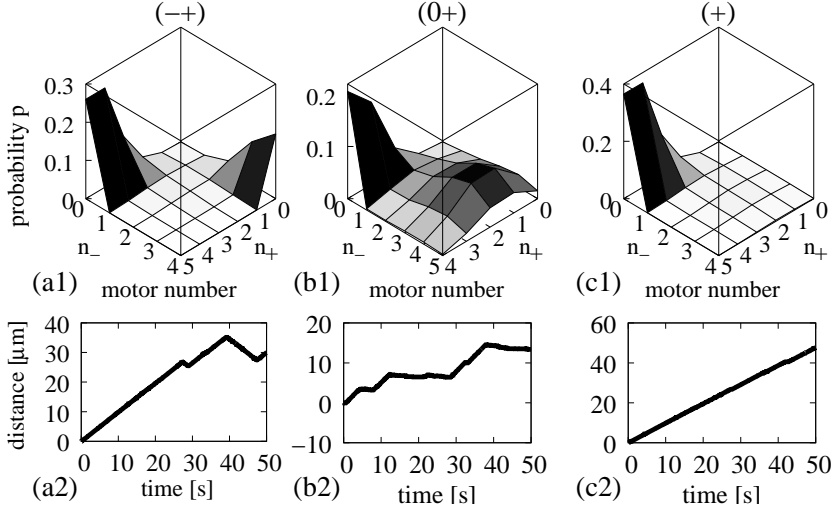}
\caption{\label{FigMotCharKinDyn}%
Motility states for the kinesin-dynein tug-of-war with parameters as given in \Tab{KinesinParameters}.
(a) A cargo which is carried by $\Np=5$ kinesins and $\Nm=4$ 'strong' dyneins is in the \PM\ motility state. The motor number probability $p(\np,\nm)$ shown in (a1) exhibits two maxima at $(5,0)$ and $(0,4)$, corresponding to fast plus and fast minus motion, respectively. As the plus motion maximum at $(5,0)$ is higher than the minus motion maximum at $(0,4)$, plus motion is more probable, and the trajectory in (a2) is biased towards the plus direction.
(b) A cargo which is carried by $\Np=4$ kinesins and $\Nm=5$ 'weak' dyneins is in the \PN\ motility state. The motor number probability $p(\np,\nm)$ shown in (b1) has two maxima, one at the fast plus motion state $(4,0)$ and one at the no-motion state $(1,4)$. The trajectory in (b2) exhibits plus motion with pauses.
(c) A cargo which is carried by $\Np=5$ kinesins and $\Nm=4$ 'weak' dyneins is in the \P\ motility state. The motor number probability $p(\np,\nm)$ shown in (c1) has one maximum at $(5,0)$, which corresponds to fast plus motion, as can be seen in the trajectory in (c2).
}
\end{figure}

Switching between fast plus and minus motors as in the \PM\ motility state has also been observed in a motility assay with only one type of non-processive motors, which move to the plus and minus end with equal probability \cite{EndowHiguchi00}. This is formally equivalent to two equal teams of non-processive unidirectional motors, with the addition that motors can interchange between being a plus and a minus motor. This situation has been investigated theoretically in the framework of two-state ratchet models, and also leads to bidirectional motion with bimodal velocity distributions \cite{BadoualProst02}. However, because of the non-processivity of the motors, a minimal number of 5 motors is required to produce bidirectional motion in this model, and a regime \PMN\ with plus and minus motion and pauses is not observed. This is in qualitative agreement with our results for unprocessive motors, \ie for high desorption constants $K$. First, a minimum number of motors $N>K$ is required in order to be in the \PM\ motility state. Second, the region of the \PMN\ motility states does not extend into the region of high desorption constants $K$, see \Fig{FigMotDiagSymN}.

\begin{figure}\centering
\includegraphics[width=11cm]{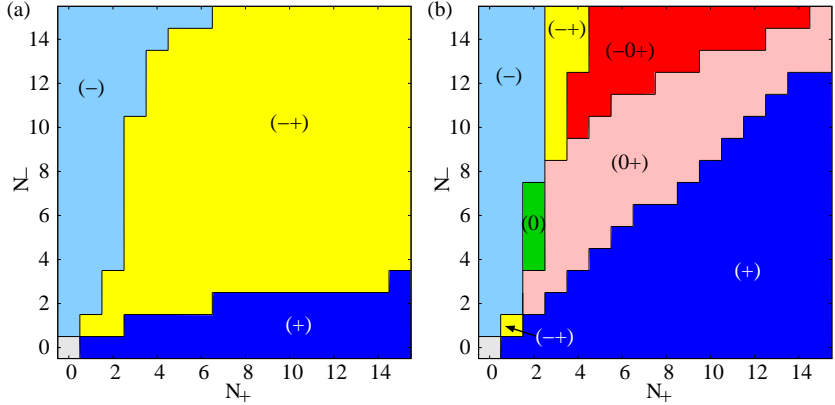}
\caption{\label{FigMotDiagKinDyn}%
Motility diagrams for the asymmetric tug-of-war of $\Np$ kinesins and $\Nm$ (a) 'strong' and (b) 'weak' dyneins, with parameters as given in \Tab{KinesinParameters}.
The colors indicate the motility states of fast plus motion \P\ (dark blue), fast minus motion \M\ (light blue), fast bidirectional motion \PM\ (yellow), fast bidirectional motion with pauses \PMN\ (red), fast plus motion with pauses \PN\ (pink) and no-motion \N\ (green). 
The motor number probability and trajectory for special choices of $\Np$ and $\Nm$ are shown in \Fig{FigMotCharKinDyn}.}
\end{figure}

\subsection{Asymmetric tug-of-war}

Bidirectional cargo transport \invivo\ is typically dependent on two different motor species for plus and minus motion. This plus-minus asymmetry can lead to net transport of the cargo in one direction. For example, in the motility state \PM, the plus motion maximum of the motor number probability can be larger than the minus motion maximum, which leads to longer plus runs compared to minus runs and to net plus motion of the cargo, see \Fig{FigMotCharKinDyn}(a). As cargo motion is no longer symmetric with respect to plus and minus motion, seven motility states are now possible, corresponding to the seven different combinations \P, \M, \N, \PM, \PN, \NM\ and \PMN\ of the maxima \P, \M, and \N. In the motility state \N, the cargo exhibits almost no motion. The motility states \PN\ and \P\ correspond to fast plus motion with and without pauses, respectively, see \Fig{FigMotCharKinDyn}(b) and (c). Analogously, in the motility states \M\ and \NM, the cargo exhibits fast minus motion without and with pauses, respectively. The motility states \PM\ and \PMN\ describe fast bidirectional motion without and with pauses, respectively. 

\Fig{FigMotDiagKinDyn} shows the motility states for the tug-of-war of $\Np$ kinesins and $\Nm$ dyneins for varying motor numbers $\Np$ and $\Nm$. As two different values of the stall force have been reported for dynein, see \Tab{KinesinParameters}, we show the tug-of-war of kinesins and 'strong' dyneins with 7 pN stall force in \Fig{FigMotDiagKinDyn}(a) and the tug-of-war of kinesins and 'weak' dyneins with 1.1 pN stall force in \Fig{FigMotDiagKinDyn}(b). 
In the tug-of-war of kinesin and strong dynein, the opposing motors are of similar 'strength', i.e. have similar stall and detachment forces and similar desorption constants. In addition, both motors are 'strong' with large stall to detachment force ratios. This is similar to the symmetric tug-of-war with strong motors, so that the cargo is mostly in the \PM\ motility state of fast bidirectional motion. Only when one motor type is much more abundant than the other, this motor types dominates the motion and the cargo exhibits fast motion into the direction of this motor type. 

In the tug-of-war of $\Np$ kinesins and $\Nm$ 'weak' dyneins, the dyneins have a much lower stall force than the kinesins. Therefore, the kinesins win the tug-of-war for dynein motor numbers smaller or of similar magnitude as the kinesin motor number, and the cargo is in the plus motion motility state  \P. If the number of dyneins is increased, the dyneins act as a brake for the kinesins, and the cargo is in the motility states \PN\ of plus motion with pauses or in the no-motion motility state \N. If the number of dyneins is increased further, the cargo is in the motility states \PM\ or \PMN\ of fast plus and fast minus motion without and with pauses, respectively. For very large number of dyneins, the cargo is in the minus motion motility state \M. The motility state \NM\ of fast minus motion with pauses does not appear, but is possible for motors with different parameters \cite{MullerLipowsky08TugOfWar}.

\begin{figure}\centering
\includegraphics[width=10cm]{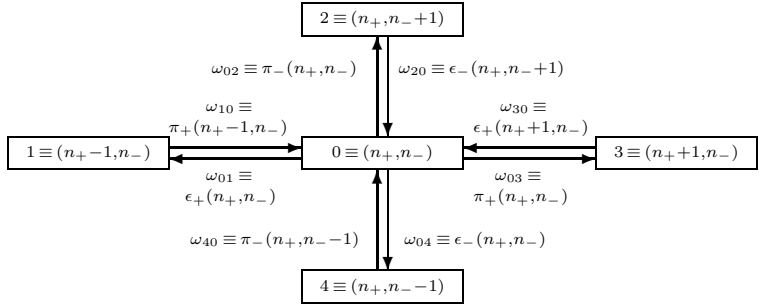}
\caption{\label{SharpMaxNeighbors}%
Relabeling of the vertex $(\np,\nm)$, its four nearest neighbors, and the transition rates connecting these five vertices. 
}
\end{figure}

\Fig{FigMotDiagKinDyn} shows that for biological motor numbers of a few motors of each type, the kinesin-dynein tug-of-war can lead to fast plus motion, fast minus motion, or fast motion in both directions. The type of motion can be regulated by choosing the appropriate number of motors on the cargo. 
Many \invivo\ cargos, which are often carried by kinesin and dynein, have indeed been observed to exhibit uni- or bidirectional motion, and also often exhibit pauses. Our results for the kinesin-dynein tug-of-war could be directly tested in \invitro\ experiments.  

In \Figs{FigMotDiagSymN}{FigMotDiagKinDyn}, the \PMN\ motility state of fast bidirectional motion with pauses appears only if the number of motors $\Np+\Nm$ is larger than 3. This can be understood by considering the state space network of \Fig{FigCartoonStateSpace}(b). If the number of motors $\Np+\Nm$ is smaller than 4, there is simply not enough room for three maxima.
This leads to an interesting prediction on the number of motors involved in transport. If there are no other causes for pausing, \eg physical obstacles, except the molecular motor tug-of-war, bidirectional cargo transport with pauses must involve $\Np+\Nm\geq4$ motors. Therefore, the observation of pauses leads to a lower bound on the motors involved in the cargo transport.

\section{Sharp maxima approximation}\label{SecSharpMax}

Although our tug-of-war model is rather simple, the  motility diagrams of \Figs{FigMotDiagSymN}{FigMotDiagKinDyn} have a complex structure. 
The qualitative aspects of these motility diagrams can be reproduced within a mean field theory for the dynamics of the average numbers of active plus and minus motors, as described in \cite{MuellerThesis}.  In particular, the mean field calculation reproduces all motility states of the stochastic tug-of-war model, with the transitions between the motility states becoming saddle-node and transcritical bifurcations. However, it does not lead to analytical expressions for the transition lines, and the quantitative agreement with the results of the stochastic calculation is poor. 
In this section, we discuss a rather simple and intuitive approximation that leads to analytical expressions for the transition lines of the motility diagram.

\begin{figure}\centering
\includegraphics[width=12cm]{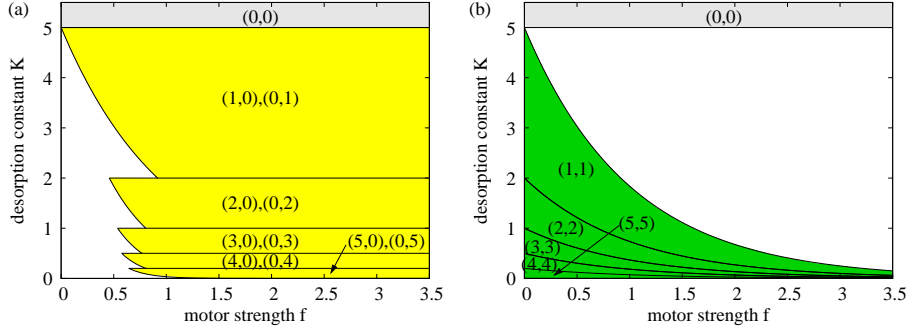}
\caption{\label{FigMotDiagSharpMaxApproxConstruct}%
Constructing the motility diagram for the symmetric tug-of-war of $\Np=\Nm=5$ plus against minus motors in the sharp maxima approximation:
The overlay of the transition lines for (a) the plus and minus motion maxima at $(n,0)$ and $(0,n)$ according to \Eq{EqSharpMaxBound} and for (b) the zero motion maxima at $(n,n)$ according to \Eq{EqSharpMaxDiag} forms the full motility diagram of \Fig{FigMotDiagSharpMax}.
The cargo is unbound (grey) for $K>N$, see \Eq{EqSharpMaxUnbound}.}
\end{figure}

We have characterized the motility states of the cargo by the maxima of the motor number probability $p(\np,\nm)$ for $\np$ active plus and $\nm$ active minus motors, because the cargo spends most of its time in such extremal states. We now focus on situations in which the probability is concentrated at the maxima. 
Thus, we now assume that the probability $p(\np,\nm)$ is non-zero only for the maxima locations and for the nearest neighbors of these points. For example, the state $(\np,\nm)$ with $0<\np<\Np$, $0<\nm<\Nm$ has the four nearest neighbors shown in \Fig{SharpMaxNeighbors}. The transition rates between these states are given by the rates for (un)binding of one plus or minus motor in \Eqs{EqCargoEpsP}{EqCargoPiP} and the analogous equations for the minus motors.
For notational simplicity, we label the extremal state by $0$ and the four neighbors by 1, 2, 3, and 4, and denote the transition rate connecting state $i$ with state $j$ by $\om_{ij}$, see \Fig{SharpMaxNeighbors}. The Master equation  then reads
\begin{eqnarray}\label{EqSharpMax1}
 	\frac{\partial}{\partial t} P_0 &=& \sum_{i=1}^4 \left( P_i\,\om_{i0} - P_0\,\om_{0i} \right),
\end{eqnarray}
for the extremal state $j=0\equiv(\np,\nm)$, and 
\begin{eqnarray}\label{EqSharpMax2}
	\frac{\partial}{\partial t} P_i &=& - P_i\,\om_{i0} + P_0\,\om_{0i}\quad\quad\quad\mbox{ with } i=1,\ldots, 4
\end{eqnarray}
for the four neighboring states. 
Note that the latter equation contains the 'sharp maxima approximation', since it ignores all neighbors of the states $i=1$, 2, 3 and 4 apart from state $0$.
The steady state solution of equations (\ref{EqSharpMax1}) and (\ref{EqSharpMax2})
fulfills the detailed balance relation $\om_{i0} P_i = P_0\om_{0i}$. The condition for the point $j=0$ to be a maximum is
\begin{eqnarray}\label{eqSharpMaxGen}
	P_0>P_i,\quad\mbox{ \ie }\quad\om_{0i} < \om_{i0}\quad\quad \mbox{ for } i=1,\ldots 4.
\end{eqnarray}
This is intuitive, stating that the rates $\om_{i0}$ leading into the maximum state $0$ should be larger than the corresponding outgoing rates $\om_{0i}$. In terms of the original transition rates as given by \Eqs{EqCargoEpsP}{EqCargoPiP}, the inequalities (\ref{eqSharpMaxGen}) lead to the conditions 
\begin{eqnarray}\label{EqSharpMaxConditionsGeneral}
\begin{array}{rcl@{\hspace{1cm}}l}
	\piP(\np,\nm) &<& \epsP(\np+1,\nm) &\mbox{ for } \np<\Np,\\
	\piM(\np,\nm) &<& \epsM(\np,\nm+1) &\mbox{ for } \nm<\Nm,\\
	\epsP(\np,\nm) &<& \piP(\np-1,\nm) &\mbox{ for } \np>0,\;\mbox{ and}\\ 
	\epsM(\np,\nm) &<& \piM(\np,\nm-1) &\mbox{ for } \nm>0
\end{array}
\end{eqnarray}
for the eight transition rates between the point $(\np,\nm)$ and its four nearest neighbors. When  the transition rates (\ref{EqCargoEpsP}) and (\ref{EqCargoPiP}) are inserted, the inequalities (\ref{EqSharpMaxConditionsGeneral}) become:
\begin{eqnarray}\label{EqSharpMaxConditionsAsym}
\begin{array}{l}
\frac{\Np-\np}{\np+1}\exp\left[-\frac{\FC(\np+1,\nm)}{(\np+1)\FdP}\right] 
	< \KP <
	\frac{\Np-\np+1}{\np}\exp\left[-\frac{\FC(\np,\nm)}{\np\FdP}\right],\;\mbox{ and}\vspace*{0.2em}\\
\frac{\Nm-\nm}{\nm+1}\exp\left[-\frac{\FC(\np,\nm+1)}{(\nm+1)\FdM}\right] 
	< \KM <
	\frac{\Nm-\nm+1}{\nm}\exp\left[-\frac{\FC(\np,\nm)}{\nm\FdM}\right].
\end{array}
\end{eqnarray}

\begin{figure}\centering
	\includegraphics[width=0.7\textwidth]{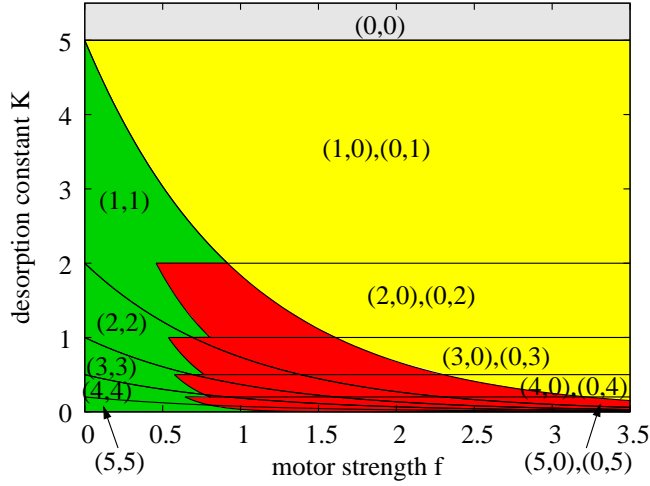}
\caption{\label{FigMotDiagSharpMax}%
The motility diagram for the symmetric tug-of-war of $N=\Np=\Nm=5$ kinesin-like plus and minus motors, calculated with the sharp maxima approximation. 
This diagram is an overlay of the two diagrams of \Fig{FigMotDiagSharpMaxApproxConstruct}, i.e. the lines separating the regions with different locations or numbers of maxima are given by equations (\ref{EqSharpMaxBound})-(\ref{EqSharpMaxUnbound}).
The colors separate different motility states as in the full numeric motility diagram of \Fig{FigMotDiagSymN5} to which it is very similar.
Parameters are as in \Fig{FigMotDiagSymN5}, except for the superstall velocity amplitude $\vB$, which is set to zero.
}
\end{figure}

Biological motors often have a small backward velocity. For example, kinesin-1 has $\vB=6\,\nanom/\s$, which is small compared to the forward velocity $\vF=1\,\mum/\s$, see \Tab{KinesinParameters}. In the limit of vanishing backward velocity $\vB=0$, the cargo force becomes
\begin{eqnarray}\label{EqForceZeroBackward}
	\FC(\np,\nm) = \left\{\begin{array}{ll}
		\max\left\{\np\FsP,\nm\FsM\right\}  	& \mbox{ for both } \np,\,\nm>0\\
		0								& \mbox{ for } \np=0 \mbox{ or } \nm=0
	\end{array}\right.  	\label{EqFCvB0}
\end{eqnarray}
and the cargo velocity is equal to the single plus (minus) motor velocity $\vC=\vFP$ (or $\vC=\vFM$) if $\np>0$, $\nm=0$ ($\np=0$, $\nm>0$), and equal to $\vC=0$ in all other cases. 
By using the relation (\ref{EqFCvB0}) in (\ref{EqSharpMaxConditionsAsym}), we obtain four simple inequalities. If these inequalities are fulfilled for a given set of motor parameters and numbers, then the state $(\np,\nm)$ is a maximum state of the steady state motor number probability $p(\np,\nm)$. In this way we can obtain all maxima for a given set of motor parameters and thereby construct the motility diagram. 

We now illustrate this for the symmetric tug-of-war of $N=\Np=\Nm$ motors with identical parameters apart from their forward direction.
In the symmetric tug-of-war, the relevant dimensionless motor parameters are given in \Eq{EqSymNondimParas}. With these parameters, one obtains from the inequalities (\ref{EqSharpMaxConditionsAsym}): 
\begin{itemize}
\item[(i)] fast motion maxima at $(\np,\nm)=(n,0)$ and $(0,n)$ 
	for
		\begin{eqnarray}\label{EqSharpMaxBound}
		\max\left(\frac{N-n}{n+1},\; N\etxt^{-n f}\right) < K < \frac{N-n+1}{n}\quad\quad\mbox{ with } n=1,\ldots,N.
		\end{eqnarray}
	The corresponding transition lines separating regions with different maxima locations are shown in \Fig{FigMotDiagSharpMaxApproxConstruct}(a). 
\item[(ii)] a no-motion maximum at $(n,n)$ for
	\begin{eqnarray}\label{EqSharpMaxDiag}
		\frac{N-n}{n+1}\,\etxt^{-f} < K < \frac{N-n+1}{n}\,\etxt^{-f} \quad\quad\mbox{ with } n=1,\ldots,N.
	\end{eqnarray}
	The corresponding transition lines separating regions with different maxima locations are shown in \Fig{FigMotDiagSharpMaxApproxConstruct}(b).
\item[(iii)] a maximum at $(0,0)$, corresponding to an unbound cargo, for
	\begin{eqnarray}\label{EqSharpMaxUnbound}
		K>N.
	\end{eqnarray}
\end{itemize}
For all other points $(\np,\nm)$ with $\np\neq\nm$ and $\np,\,\nm>0$, the inequalities (\ref{EqSharpMaxConditionsAsym}) have no solution. This means that all maxima that are obtained within the numerical calculation are also found in the sharp maxima approximation, and vice versa.
Drawing the transition lines between the different maxima regions in the $(f,K)$-plane leads to the motility diagram shown in \Fig{FigMotDiagSharpMax}. The maxima $(n,0)$ and $(0,n)$  with only one motor type bound to the filament define the \PM\ regime (yellow), while the maxima $(n,n)$ with both types of motors bound define the \N\ regime (green). The overlap region is the \PMN\ regime (red). This sharp-maxima motility diagram is in good agreement with the exact motility diagram shown in \Fig{FigMotDiagSymN5}. The rather simple analytic equations of the sharp maxima approximation can therefore be used to estimate the transition lines. 
The sharp maxima approximation tends, however, to overestimate the existence region of a maximum, and therefore the  \PMN\ region, because the probability for a given vertex is only compared to the probabilities of its next neighbors. 

\begin{figure}\centering
\includegraphics[width=11cm]{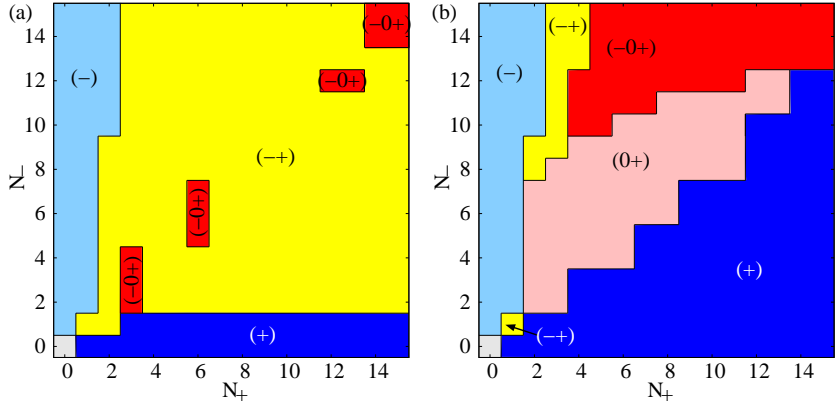}
\caption{\label{FigMotDiagKinDynSharpMax}%
Motility diagrams for the kinesin-dynein tug-of-war, calculated with the sharp maxima approximation. 
The motor parameters are for kinesin-1 and (a) strong dynein or (b) weak dynein as given in \Tab{KinesinParameters}, except for the superstall velocity amplitude $\vB$, which is set to zero. The motility diagrams from the sharp maxima approximation agree well with the exact motility diagrams shown in \Fig{FigMotDiagKinDyn}.}
\end{figure}

We now apply the sharp maxima approximation to the kinesin-dynein tug-of-war. We fix the motor parameters to the values for kinesin-1 and strong or weak dynein as given in \Tab{KinesinParameters}. For $\Np$ kinesins and $\Nm$ dyneins, a state $(\np,\nm)$ with $0\leq\np\leq\Np$ and $0\leq\nm\leq\Nm$ is counted as a maximum of the motor number probability if all four inequalities (\ref{EqSharpMaxConditionsAsym}) are fulfilled. With this method, the number and locations of the maxima are determined, which defines the motility state for these values of $\Np$ and $\Nm$. When $\Np$ and $\Nm$ are varied, the motility diagrams shown in \Fig{FigMotDiagKinDynSharpMax} are obtained.
They agree well with the exact motility diagrams shown in \Fig{FigMotDiagKinDyn}. Again, the sharp maxima approximation tends to overestimate the existence region of a maximum, and therefore especially the \PMN\ region.

The sharp maxima approximation is based on the simple idea that the probability is highly concentrated around single cargo states $(\np,\nm)$. As described in this section, we have taken into account only the four nearest neighbors of the center states. It is possible to systematically expand this scheme to the 13 next-nearest neighbors etc. However, even including only the eight diagonal neighbors $(\np\pm 1,\nm\pm 1)$ already leads to analytically intractable expressions since the state network then contains cycles. 
In addition, this more elaborate approximation scheme lead only to relatively small improvements compared to the nearest-neighbor approximation,
see \cite{MuellerThesis}. Therefore, we focused here on this latter approximation, which quantitatively describes the transition lines of the motility diagrams by four simple inequalities (\ref{EqSharpMaxConditionsAsym}), compare \Fig{FigMotDiagSharpMax} and \Fig{FigMotDiagSymN5} as well as \Fig{FigMotDiagKinDynSharpMax} and \Fig{FigMotDiagKinDyn}. These inequalities can thus be used to estimate the motility state of a given experimental system without having to resort to numerical calculations or simulations. 
Furthermore the inequalities (\ref{EqSharpMaxConditionsGeneral}) show that a transition between motility states occurs when the rates for binding and unbinding of a single motor in a maximum state $(\np,\nm)$ become comparable. 

\section{Summary and discussion}

We have studied a stochastic tug-of-war model for bidirectional cargo transport by two antagonistic teams of molecular motors. In this model, the motors act independently and are coupled only via the mechanical interaction with their common cargo.  

A particularly instructive case is the symmetric tug-of-war of plus and minus motors with the same parameters except their forward direction. In this symmetric case of 'equally strong' opposing motors, the cargo can be in one of three 'motility states', depending on the values of the motor parameters: For weak motors with low stall to detachment force ratio, the cargo is in the no-motion motility state \N\ and exhibits only slow motion. For strong motors with high stall to detachment force ratio, on the other hand, the cargo is in the motility state \PM\ characterized by switching between fast plus and fast minus motion. In the intermediate case, it exhibits the \PMN\ motility state of fast bidirectional motion with interspersed pauses. The latter two motility states correspond to experimentally observed bidirectional cargo motion and have been previously attributed to a coordination mechanism rather than to a tug-of-war. The fast bidirectional motion is obtained in our tug-of-war model via a dynamic instability arising from the nonlinear force-dependence of the single-motor unbinding rate. This instability leads to unbinding cascades of one type of motors, so that there is a high probability of having only the opposing motor type active at a given time. 

In the asymmetric tug-of-war, \eg between kinesins and dyneins, the seven motility states \P, \M, \N, \PM, \PN, \NM, and \PMN\ are possible, which correspond to all possible combinations of fast plus motion \P, fast minus motion \M\ and pausing \N. 

We have characterized the parameter dependence of the cargo motility in terms of the 'motility diagrams', which, similar to phase diagrams, describe how the different motility states depend on the system parameters.
For the symmetric tug-of-war, the relevant parameters are the stall to detachment force ratio $f$, and the desorption constant $K$. The transition lines separating the motility states \N, \PM\ and \PMN\ lie in biologically relevant parameter ranges of both $f$ and $K$. This means that the cell can use  fine-tuning of these parameters in order to achieve large changes of the cargo motility, \ie in order to regulate its cargo traffic. 
For the kinesin-dynein tug-of-war, we have varied the numbers of kinesins and dyneins. For biologically relevant numbers of a few kinesins and dyneins, the cargo mostly exhibits fast plus motion, fast minus motion, or fast bidirectional motion. However, even in the latter case, the cargo may cover a net travel distance, \eg because runs in the plus direction may be longer and/or faster than runs in the minus direction. The net direction and the speed of cargo motion can therefore be regulated in two ways: by choosing the appropriate number of motors on the cargo, and by tuning the motor properties.

The motility state \PMN\ of fast bidirectional motion with interspersed pauses only appears when the total number of plus and motors, $\Np+\Nm$, becomes larger than 3. If there are no other reasons for pausing, \eg physical obstacles, this leads to the prediction that bidirectional cargo transport which exhibits pauses is carried out by at least 4 motors. 

The transition lines separating the different motility regions in the motility diagrams have complex shapes, which are hard to understand at first sight. We have  used a 'sharp maxima' approximation to obtain a simple and intuitive derivation of these transition lines. This approximation explains the transitions between the motility states as determined by the balance of the motor binding/unbinding dynamics under force. The sharp maxima approximation reproduces the complex tug-of-war motility diagrams quantitatively by using only four simple inequalities as given by  (\ref{EqSharpMaxConditionsAsym}). It can therefore be used to determine the motility state of a given experimental system without having to do simulations or numerical calculations. 

The understanding gained from our model allows us to speculate on constraints imposed on the numbers and properties of the motors involved in cellular bidirectional transport. In order to obtain fast bidirectional cargo motion, the motors must produce large forces - compared to their detachment forces - in order to initiate the unbinding cascade which leads to this fast motion. This dynamic instability is necessary to achieve effective motor cooperation without the need for an additional coordination complex. This may be one reason for the counterintuitive property of kinesin-1 to be able to produce high forces, corresponding to a high stall force, without being able to sustain them for a long time, as reflected in the lower detachment force. The dynamic instability also provides an explanation for kinesin's intermediate processivity of 'only' about a second before unbinding: longer unbinding times would slow down the unbinding cascade and lead to lower velocities and longer pauses. 
In order for cargo motion to be easily controllable, a small change in the motor parameters should lead to a large change in cargo motility. Such a response behavior is obtained in our model when the motor binding and unbinding rates are of similar order of magnitude corresponding to a desorption constant of the order one.

Finally, why should the cell use two teams of opposing motors on a single cargo at all? Using only unidirectional motors on a single cargo is not sufficient to maintain cellular transport in an efficient way. As the cellular microtubule cytoskeleton is typically isopolar, plus end cargos would accumulate at the microtubule plus ends, and minus end cargos at the microtubule minus ends. Even if motors are interchanged at the microtubule ends, which would solve the cargo jamming problem, it would still leave the problem how the plus (minus) motors reach the minus (plus) ends in order to start their journeys. Having motors of opposite directionality on the same cargo is an elegant solution to this problem. 
In addition, bidirectional cargos can easily be regulated. By using the sensitivity of the cargo motility to the motor properties, the cell could influence the cargo motion easily by fine-tuning the parameters of the two motor species.


\begin{acknowledgements}
SK was supported by Deutsche Forschungsgemeinschaft (Grants KL818/1-1 and 1-2) and by the National Science Foundation through the Physics Frontiers Center-sponsored Center for Theoretical Biological Physics (Grant PHY-0822283).
\end{acknowledgements}

\appendix

\section[\hspace*{2cm}Cargo unbinding]{Cargo unbinding}\label{AppCargoUnbinding}

In this appendix, we show that the bound motion of the cargo is independent of the rates for cargo unbinding and binding. 

\begin{figure}\centering
\includegraphics[width=0.35\textwidth]{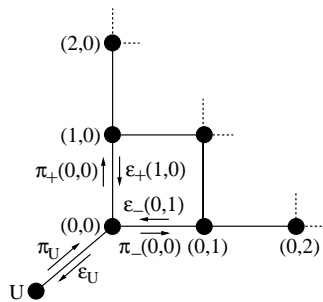}
\caption{\label{FigNetUnbinding}%
Lower left corner of the tug-of-war states network of \Fig{FigCartoonStateSpace}(b), with the addition of state $U$ representing an unbound freely diffusing cargo.}
\end{figure}

We have characterized the cargo state  by the numbers $\np$ of plus and $\nm$ of minus motors that link it to the filament. Cargo unbinding can only occur when no motors link the cargo to the filament, \ie from the state $\np=\nm=0$. This state can be split into two states, see \Fig{FigNetUnbinding}: an unbound diffusive state $U$ where the cargo is far away from the filament, and a state $(0,0)$ where the cargo is close to the filament as if bound by motors (but no motor \textit{is} bound). The cargo is in the $(0,0)$ state, if for example the last motor linking it to the filament has just unbound a short time ago. From state $(0,0)$, the cargo can 'rebind' to the filament because of binding of a plus or a minus motor. Alternatively, the cargo can 'unbind' from the filament by diffusing away into the unbound state $U$. This is described by the unbinding rate $\epsU$. 
The binding rate $\piU$ from state $U$ to state $0$ describes the probability that the freely diffusing cargo comes close enough to the filament for a motor to bind. The rates $\piU$ and $\epsU$ relate to diffusion of the cargo in the surrounding solution and therefore depend on the geometry of the system, \eg the location of the filaments or the volume and viscosity of the surrounding medium. 

The rates leading from the $(0,0)$ state to the states $(1,0)$ and $(0,1)$ with one bound plus \resp one bound minus motor are the rates for the binding of the respective motor when the cargo is close to the filament.  These are the rates $\piP(0,0)=N\piOP$ \resp $\piM(0,0)=N\piOM$ used in \Sec{SecModel}.
Actually, all the rates $\piPM(\np,\nm)$ use the single motor rates $\piOPM$ which are the rates for the binding of a plus \resp minus motor to the filament when the cargo is close to the filament. 

Let $P(\np,\nm)$ denote the steady state probability of $\np$ plus and $\nm$ minus motors linking the cargo to the filament, and $P(U)$ the steady state probability of being in the diffusive unbound state $U$. As experiments only monitor bound cargos, we consider conditional probabilities, conditioned on the cargo being bound:
\begin{eqnarray}
	p(\np,\nm) \equiv  P(\np,\nm) / \left[1-p(U)\right]  
\end{eqnarray}
for $0\leq\np\leq\Np$ and $0\leq\nm\leq\Nm$. The state $(0,0)$ is included here, since in this state the cargo has not diffused away yet and appears as bound to an experimenter.

We now show that in the steady state, these conditional probabilities do not depend on the rates $\piU$ and $\epsU$. In order to do this we consider the calculation of the steady state via the diagram or Kirchhoff method. This method is similar to the Kirchhoff rules for electrical networks and is reviewed in \cite{Schnakenberg76}.
First, one constructs the complete set of partial diagrams for the state space network. A partial diagram is a diagram with the maximum number of lines that can be included without forming a closed cycle. From these partial diagrams, one constructs the directional diagrams for each state $n$ of the network: add arrows to each partial diagram flowing towards state $n$. In our case, $n$ can be one of the states $(\np,\nm)$ with $0\leq\np\leq\Np$ and $0\leq\nm\leq\Nm$, or $U$. Each directional diagram represents the product of the rates of the arrows of the directional diagram. 
Let $\Pi_i^n$ denote the product of rates associated with the $i$th directional diagram of state $n$. Then the steady state probability of state $n$ is given by
\begin{eqnarray}
	P(n)  = {\textstyle\sum_i^n}\,\Pi_i^n / S.
\end{eqnarray}
Here, $S = \sum_n\sum_i^n\Pi_i^n$ is the sum of all directional diagrams. $\sum_i^n$ denotes the sum over all directional diagrams $i$ of state $n$, and $\sum_n$ the sum over all states. Because of the equality
\begin{eqnarray}
	1 - P(U)  = \left(S - {\textstyle\sum_i^U}\Pi_i^U\right) / S,
\end{eqnarray}
the conditional probability to be in state $n$ under the condition not to be in state $U$ is
\begin{eqnarray}\label{eq:Unbind_StatProb}
	p(n) = \frac{P(n)}{1 - P(U)}  = \frac{{\textstyle \sum_i^n}\Pi_i^n}{S - \sum_i\Pi_i^U} \quad\mbox{ for } n\neq U.
\end{eqnarray}
Now consider the directional diagrams $\Pi_i^n$ in more detail. They all contain the line $U-(0,0)$, as this line is never part of a cycle. Any directional diagram $\Pi_i^U$ for state $U$ contains the rate $\eps_U$, but not the rate $\pi_U$, as all paths in these directional diagrams must end in state $U$. On the other hand, all directional diagrams   $\Pi_i^n$ with $n\neq U$ contain the rate $\pi_U$ but not $\eps_U$ (otherwise there would be a path ending in $U$, but all paths must end in $n$). Thus in \Eq{eq:Unbind_StatProb}, $\eps_U$ does not appear (the terms in the total sum $S$ in the denominator containing $\eps_U$ are substracted) and the rates $\pi_U$ appear exactly once in every term in the numerator and denominator and thus cancel out. Therefore, $p(n)$ does not depend on the rates $\pi_U$ and $\eps_U$.

\bibliographystyle{spmpsci}
\bibliography{/usr/home/mmueller/TeXFiles/Bibliographies/Ref/Axon,%
/usr/home/mmueller/TeXFiles/Bibliographies/Ref/BidirMotorTransport,%
/usr/home/mmueller/TeXFiles/Bibliographies/Ref/MotorRunExperimentsRateTheory,%
/usr/home/mmueller/TeXFiles/Bibliographies/Ref/MotorRegulation,%
/usr/home/mmueller/TeXFiles/Bibliographies/Ref/Biology,%
/usr/home/mmueller/TeXFiles/Bibliographies/Ref/CellTrafficTheory,%
/usr/home/mmueller/TeXFiles/Bibliographies/Ref/CollectiveMotors,%
/usr/home/mmueller/TeXFiles/Bibliographies/Ref/ComputerSim,%
/usr/home/mmueller/TeXFiles/Bibliographies/Ref/Misc,%
/usr/home/mmueller/TeXFiles/Bibliographies/Ref/MotorTheory,%
/usr/home/mmueller/TeXFiles/Bibliographies/Ref/MotorReview,%
/usr/home/mmueller/TeXFiles/Bibliographies/Ref/Noneq,%
/usr/home/mmueller/TeXFiles/Bibliographies/Ref/Kinesin,%
/usr/home/mmueller/TeXFiles/Bibliographies/Ref/Dynein,%
/usr/home/mmueller/TeXFiles/Bibliographies/Ref/Myosin,%
/usr/home/mmueller/TeXFiles/Bibliographies/Ref/MPIKG,%
/usr/home/mmueller/TeXFiles/Bibliographies/Ref/UnconventionalKinesin}

\end{document}